\documentclass[prl,twocolumns,floatfix,showpacs,nofootinbib]{revtex4}
\usepackage[T2A]{fontenc}
\usepackage[cp1251]{inputenc}
\usepackage{graphicx}
\usepackage{epsf}
\usepackage{epsfig}
\usepackage{amssymb,amsthm,amsmath}
\usepackage{latexsym}

\begin{document}

\preprint{}
\title{
Spatial image of reaction area from scattering. I:\\
{\small $SO_{\mu}(2.1)$ algebra formalism.}}

\author{M. V. Polyakov}
 \email{Maxim.Polyakov@tp2.ruhr-uni-bochum.de}
  \affiliation{Institut f\"ur Theoretische Physik II
   Ruhr-Universitaet Bochum, NB6 D-44780 Bochum, Germany}

\author{O.N. Soldatenko, A.N.Vall}
 \email{vall@irk.ru}
  \affiliation{Department of Theoretical Physics, Irkutsk State University, Irkutsk, 664003
  Russia}
\author{A.A.Vladimirov}
 \email{avlad@theor.jinr.ru}
  \affiliation{
  Bogoliubov Laboratory of Theoretical Physics,
  JINR, 141980, Moscow Region, Dubna, Russia}

\keywords{Impact parameter, Eikonal approximation, hard processes}

\begin{abstract}
We develop general formalism of how to relate scattering
amplitudes for exclusive processes to spatial image of target
hadron. More precisely we show how to determine the spatial
distribution of outgoing particles in the space of so-called
nearest approach parameter. This parameter characterizes the
position in the coordinate space the place where the final
particle is produced.
\end{abstract}

\maketitle

\section{Introduction}
\fontsize{12pt}{14.5pt}\selectfont

Famous experiment by Hofstadter et al. \cite{hofstadter}
demonstrated that hadrons are not point-like particles and they
have non-trivial spatial structure. Recently interest to the
spatial images of hadrons was revived with development of the
theory of hard exclusive processes, which can be systematically
studied using formalism of Generalized Parton Distributions (GPDs)
(for reviews see \cite{reviewGPDs}). It was demonstrated
\cite{Budkard} that GPDs provide an information about spatial
distribution of partons in the transverse plane.

In the present paper we develop general formalism of how to relate
scattering amplitudes for exclusive processes to spatial image of
target hadron. More precisely we show how to determine the spatial
distribution of outgoing particles in the space of so-called
nearest approach parameter. This parameter characterizes the
position in the coordinate space the place where the final
particle is produced.

One-particle states of a free relativistic particle are defined by
the 10-parametric group of motion of Minkowski space, which is
given by generators of the Poincare group: $ M_{\mu \nu}$ and
$P_{\nu}$. States can be specified by choosing one or another
subgroup. It was found that for description of deep inelastic
processes it is useful to consider the "infinite momentum frame"
\cite{Soper1,Soper2}. We remind briefly necessary for this paper
ingredients.

The three following generators of the Poincare group:
\begin{equation}\label{p1}
  B_{1}=\frac{1}{\sqrt{2}}(K_1 + J_2 )~ ,~ B_{2}=\frac{1}{\sqrt{2}}(K_2 - J_1 )~
  ,~  J_3~,
\end{equation}
with
$$K_{i}=M_{i0}~~ ,~~ \epsilon _{ijk}J_{k}= M_{ij}~,  $$
satisfy the system of commutation relations:
\begin{equation}\label{p2}
\begin{split}
&[B_1, B_2]=0~,\\
&[J_3, B_1]=iB_2~,\\
&[J_3, B_2]=-iB_1~, \\
&[B^2 _{1}+B^2 _{2}, J_3]=0~, \\
&[p_{i},B_{j}]=i\delta _{ij}p^+~, \\
&[p^+, B_{i}]=[p^+,J_3]=0~.
\end{split}
\end{equation}
Here $p_i$ is the operator of 3-momentum, $p^+
=\frac{1}{\sqrt{2}}(p_0  + p_3)$.  The above commutation relations
imply that generators $B_1$, $B_2$ and $J_3$ form the $E(2)$
algebra. Casimir operator of this algebra is $B^2=B^2_1 + B^2_2 $
. Because of the last commutation relation, this algebra can be
realized in the three-dimension momentum space restricted by the
condition $p^+ =const$ on states $|\vec{\lambda },p^+ \rangle  $.
These states satisfy equations:
\begin{equation}\label{p3}
\begin{split}
   &B^2 |\vec{\lambda },p^+ \rangle  = \lambda ^2 |\vec{\lambda },p^+ \rangle ~,\\
   & B_1 |\vec{\lambda },p^+ \rangle = \lambda _1 |\vec{\lambda },p^+ \rangle  ~,\\
   & B_2 |\vec{\lambda },p^+ \rangle = \lambda _2  |\vec{\lambda },p^+ \rangle  ~,\\
   & \lambda ^2 = \lambda _1 ^2 + \lambda _2 ^2~.
\end{split}
\end{equation}

Operators $B_1$ \ , $B_2\ $,  $ B^2 $ \ on the $p^+ =const$
surface have complicated expression. However in the limiting case
of $ p_{\bot}/p \ll 1 $ the corresponding operators are given by
simple expression:
\begin{equation}\label{p4}
  B_1=-ip^+ \frac{\partial }{\partial p_1}~ ,~ B_2=-ip^+ \frac{\partial }{\partial
  p_2}~.
\end{equation}
Substituting these operators into equations (\ref{p3}), we obtain
the following expression for the corresponding wave function in
momentum representation:
\begin{equation}\label{p5}
  \langle \vec{p}_{\bot},p^+ |\vec{\lambda },p^+ \rangle =\exp \left( i \frac{\vec{p}_{\bot}\cdot \vec{\lambda
  }}{p^+}\right)~.
\end{equation}
Now if we introduce dimensional parameter
$$ \vec{b}=\frac{\vec{\lambda }}{p^+}~, $$  than the wave function
(\ref{p5}) has the form of the eikonal kernel in the plane of
impact parameter $\vec{b}$. It allows  us  to
 interpret in the infinite
momentum frame the eigenstates of the algebra (\ref{p2})
$|\vec{\lambda },p^+ \rangle $, as the states with fixed spatial
position in the transverse plane  \cite{Andrianov,Huszar}. In the
infinite momentum frame the transformation from the transverse
momentum to the transverse position representation is reduced to
simple two-dimensional Fourier transformation (see
eq.~(\ref{p5})). That is the transformation which relates the GPDs
to the parton distributions in the spatial transverse plane
\cite{Budkard}. Note that such interpretation is possible only in
the infinite momentum space, in order to ensure the condition
$p_\perp/p \ll 1$. For the hard exclusive processes the latter
condition corresponds to the condition $t/Q^2 \ll 1$, here $t$ is
the 4-momentum transfer squared and $Q^2$ is the hard scale for
the process. First, in experiments the ratio $t/Q^2$ can be not be
not very small and corrections to the eikonal picture can be
sizeable. Secondly, the simple two-dimensional Fourier transform
can not be used to obtain spatial distributions of partons for
hard processes with large transverse momentum transfer (like
hadron form factors, wide angle Compton scattering, etc.
\cite{wide}) Therefore our aim here is to develop general
formalism which would allow "spatial imaging" of hadrons in
arbitrary reference frame.

We consider another subalgebra of Poincare group.  This is
subalgebra of operators:
\begin{equation}\label{p6}
d_{i}=\frac{1}{p^2}M_{ij}p_{j}~.
\end{equation}
It is realized on the surface $p^2 = const$ . Below we analyze in
detail this algebra and use it for description and building one
particle states with defined quantum numbers.

 \section{Operator of the nearest approach}
\begin{figure}[h]
\begin{center}
\label{f:1}
\includegraphics*[scale=0.7] {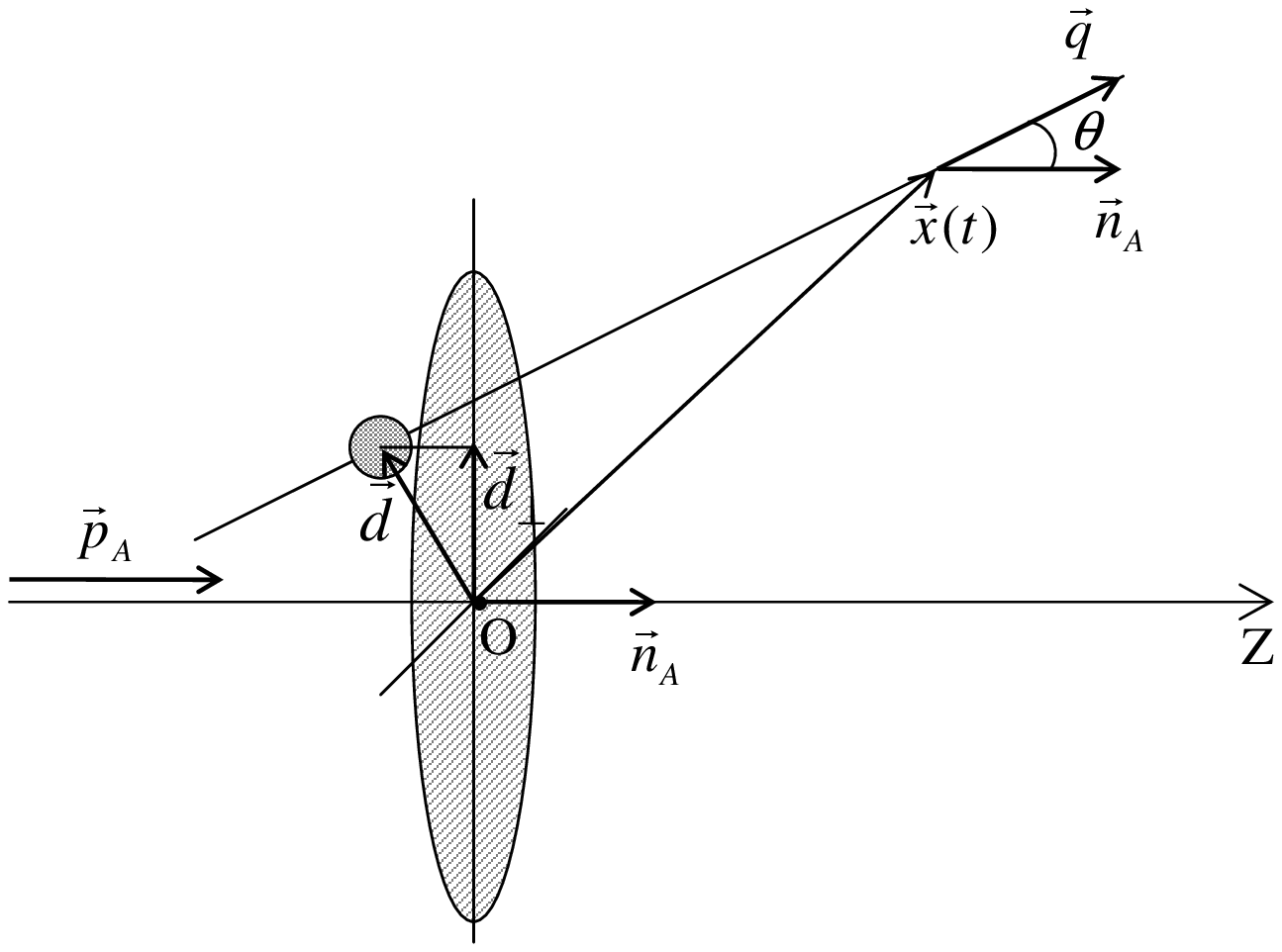}
\caption{\small A classical trajectory $\vec{X}(t)$ of an
asymptotically free particle with momentum $\vec{q}$ is proceeded
into reaction area, is characterized by a vector of maximal
approach $\vec{d}$ to the select point "O". In the labor frame
this point is a center of target, in the center mass frame it is a
meeting point of beams.}
\end{center}
\end{figure}

Let us consider a trajectory of a free spinless particle, which
moves with arbitrary initial conditions (Fig.1) \cite{Vall1}:

\begin{equation}\label{x}
  X_{i}(t)=\frac{q_{i}}{\sqrt{\vec{q}^2+m^2}}(t-t_0)+\xi _{i} ~ ,~
  i=1,2,3~,
\end{equation}
with $\vec{q}$ - momentum of the particle , $m$ -  mass of the
particle. A distance between the particle and the coordinate
origin "O"~ is
$$D(t)=(\vec{X}(t)\vec{X}(t))^\frac{1}{2}~.$$

At some moment of time $t=\tau$ the distance $D(\tau)$ has a
minimal value. The time $\tau$ is defined by the extremum
condition:
\begin{equation}\label{4}
\frac{d}{dt}D(t)\Big|_{t=\tau}=\frac{1}{D(t)}\Big(X_{i}(t)
\frac{d}{dt}X_{i}(t)\Big)\Big|_{t=\tau}=0~ .
\end{equation}
Using (\ref{x})\ , we obtain:
\begin{equation}\label{tau}
\begin{split}
\tau &=t_0-\frac{(\vec u \vec \xi)}{u^2}~, \\
u_{i}&= \frac{q_{i}}{\sqrt{\vec{q}^2+m^2}}~.
\end{split}
\end{equation}
Substituting $t=\tau $ into $D(t)$, we get the following
expression for components of the vector of the nearest approach
$d_i$:
\begin{equation}\label{6}
d_i=X_{i}(\tau)=-\frac{u_i}{u^2}(\vec u \vec\xi)+\xi_i ~.
\end{equation}

Let us introduce the operator of orbital momentum $L_k\ ,\
k=1,2,3$, in usual way:
\begin{equation}\label{L}
L_k=\varepsilon_{klm}\xi_l q_m ~.
\end{equation}
Then the expression (\ref{6}) can be rewritten as:
\begin{equation}\label{8}
d_i=\frac{1}{q^2}\varepsilon_{ijk}q_j L_k
=\frac{1}{q^2}M_{ij}q_{j}~ .
\end{equation}

We see that for each classical trajectory $\vec{X}(t)$ of an
asymptotically free particle, which moves out from the reaction
area with momentum $\vec{q}$, we can find the vector $\vec{d}$ of
the nearest approach of the trajectory to the point "O". We can
interpret this vector as an effective coordinate of the particle
creation region.

Note that for the process reversed in time,  the vector $\vec{d}$
corresponds to the impact parameter. The impact parameter is
useful for geometrical interpretation of scattering which is based
on the eikonal approximation for the scattering amplitude. In the
eikonal approximation the states with definite impact parameter of
the initial particle are defined as semi-classical approximation
of states with definite orbital momentum. The impact parameter
representation is a good toll for studying diffractive processes,
see e.g \cite{Predazzi}. However for the small value of impact
parameter semi-classical approach is not valid. It is especially
clearly seen from the uncertainty principle: the states with
finite momentum and infinitesimal impact parameter are forbidden.
So, there are significant quantum effects in the region of small
impact parameters. Also a calculation of matrix elements of
$\mathrm{S}$-matrix needs consistent quantum-mechanical definition
of states. In this case it is a state with a define value of the
$\vec{d}$.

\section{Quantum description of the nearest approach parameter. }

Using a standard quantization procedure based on the canonical
commutation relations:
$$[\xi _{i}\ q_{j}]=\imath \delta _{ij} , $$
we get a system of commutation relations:
\begin{equation}\label{algebra1}
\begin{split}
d_{i}&=\frac{1}{q^{2}}(\varepsilon_{ijk}q_{j}L_{k}-iq_{i})~,\\
d_{i}&=(d_{i})^{+}\ \ ,\ \ [d_{i} q^2  ]=0~, \\
 [d_{i}L_{j}]&=-i  \varepsilon_{ijk}d_{k}~,\\
[L_{i}L_{j}]&=i  \varepsilon_{ijk}L_{k}~,\\
[d_{i}d_{j}]&=\frac{-i  }{q^{2}}\varepsilon_{ijk}L_{k}~,\\
[d_{i}q_{j}]&=i  \delta_{ij}-\frac{i  }{q^{2}}q_{i}q_{j}~,\\
[d_{i},q_{j}q_{k}]&=i  \delta_{ij}q_{k}+i  \delta_{ik}q_{j}-\frac{2i  }{q^{2}}q_{i}q_{j}q_{k}~,\\
.......&.............\textrm{ etc.}
 \end{split}
\end{equation}
We see  that operators $ d_{i}$ and $L_{j}$ form algebra of
$SO(3,1)$ group on the sphere $q^2=const$. Corresponding Casimir
operator of this algebra becomes a number:
$$  \hat{C}=d^2 -\frac{1}{q^2}L^2\equiv \frac{1}{q^2}~.$$

Operators $d_{1}\ ,\ d_{2} \ ,\ L_{3}$ form a nontrivial
subalgebra:
\begin{equation}\label{algebra2}
\begin{split}
[d_1 d_2]&=-\frac{i}{q^2} L_3~, \\
[d_1 L_3]&=-i d_2~, \\
[d_2 L_3]&=i d_1~.
 \end{split}
\end{equation}
This is algebra of $SO(2,1)$ group. Its properties are
investigated and presented in details in monograph
\cite{Vilenkin2}. This group is non compact and it has three
series of unitary representations in space of quadratically
integrated functions: general, non compact and discreet. Casimir
operator of this algebra is:
\begin{equation} \label{algebra3}
\hat{K}=d^{2}_{\perp}-\frac{1}{q^2}L_3^2 \ \ ,\ \
d^{2}_{\perp}=d_1^2+d_2^2\ \ ,\ [d_{1,2}\hat{K}]=0\ ,\
[d_{3}\hat{K}]\neq 0~.
\end{equation}

We interpret the eigenvalue of this operator $\hat{K}$ in the
continuous spectrum as a quantum-mechanic generalization of a
squared effective distance of the created particle, as it was
defined above. Here we note, that relations (\ref{algebra2})
suppose a definite choice of the $z$ direction. We link this
direction with direction of the momentum of projectile particle
$\vec{p}_{A}$ .

A realization of the algebra (\ref{algebra2}) is connected closely
to the geometrical properties of the transverse momentum space -
the space on which this algebra is realized. Let us consider
unitary transformations of the momentum space generated by the
generators $d_1$ and $d_2$ \cite{Vall2}:

\begin{equation}\label{algebra4}
\begin{split}
q_{i}'&=e^{-i\vec{p}\cdot \vec{d}}q_{i}e^{i\vec{p}\cdot \vec{d}}\ \ ,\ \ i=1,2~, \\
\vec{p}=p\vec{n}\ ,\ \vec{n}&=(\cos\varphi \ ,\  \sin\varphi )\ ,\
\vec{n}\cdot \vec{d}=n_1 d_1 +n_2 d_2\ \ ,\ -\infty <p<+\infty~.
\end{split}
\end{equation}
If we differentiate this relations in respect to $p$, we  get the
following Lee equation for $q_{i}'$:
\begin{equation}\label{algebra5}
\begin{split}
\frac{dq_{i}^{'}}{dp}&=n_{i}-\frac{\vec{n}\cdot
\vec{q}_{\bot}^{'}}{q^2}q_{i}^{'}\ \ ,\ \ i=1,2~,
\\ & q_{i}^{'}(p=0)=q_{i}~.
\end{split}
\end{equation}
This equation can be easily  integrated:
\begin{equation}\label{algebra6}
q_{i}^{'}=\frac{q_{i}+ \left( q \sinh(\frac{p}{q}) +  \vec{n}\cdot
\vec{q}_{\bot}\cosh(\frac{p}{q}) -\vec{n}\cdot \vec{q}_{\bot}
\right )n_{i} }{\cosh(\frac{p}{q})+ \frac{(\vec{n}\cdot
\vec{q}_{\bot})}{q}\sinh(\frac{p}{q}) }\ \ ,\ \ i=1,2 ~.
\end{equation}
It is convenient to introduce new parameter $\vec{\kappa }$
related to the parameter $\vec{p}$ by:
\begin{equation}\label{algebra7}
\vec{\kappa }=q\vec{n}\ \tanh\left(\frac{p}{q}\right) ~,
\end{equation}
then we have:
\begin{equation} \label{q+x}
q'_i=\Big(\big[1+\frac{(\vec{\kappa}\vec{q})(1-B)}{\kappa^2}\big]\
\kappa_i+B\,q_i\Big)
\Big(1+\frac{(\vec{\kappa}\vec{q})}{q^2}\Big)^{-1}\ \equiv
\vec{q}\oplus\vec{\kappa}~,
\end{equation}
with
$$B=\sqrt{1-\frac{\kappa^2}{q^2}} ~.$$
We can see from (\ref{q+x}) , that the  $SO(2,1)$ motion of the
transverse momentum space is the nonlinear transformation. The
operation $\oplus$ is not a group operation. Because of a product
of two such operations contains not only the resulting operation
$\oplus $, but also a rotation. But the operation $\oplus $ has a
set of group properties. In particular, transformations
(\ref{q+x}) posses the invariant interval $ds$ :
\begin{equation}\label{algebra8}
\begin{split}
ds^2 &=G_{ik}\ dq_{i}dq_{k}~,\\
G_{ik}&=\frac{q^2}{q_3 ^2}\left ( \delta_{ik}  +\frac{q_{i}q_{k}}{q_3 ^2}  \right )\ \ ,\ \ i,k=1,2~, \\
q_3 ^2 &=q^2-q_{\bot}^2~.
\end{split}
\end{equation}
Ricci tensor corresponding to the metric tensor $G_{ik}$ has the
form:
\begin{equation}\label{algebra9}
R_{ik}=- \frac{1}{q_3 ^2} \left ( \delta_{ik}
+\frac{q_{i}q_{k}}{q_3 ^2}  \right )~.
\end{equation}
The corresponding scalar curvature is:
\begin{equation}\label{algebra10}
  R=-\frac{2}{q^2}~.
\end{equation}
We see that the transverse momentum plane which is obtaine by
transformations (\ref{q+x}) from fixed $\vec{q}_{\bot}$ by
changing parameter $\vec{\kappa}$ in the region $|\vec{\kappa}|^2
\leq q^2 $, is the space with constant negative curvature.

Let us now show that transformations (\ref{q+x}) does not change
the sign of $q_3$, i.e. forward ($q_3 >0$) and backward ($q_3 <0$)
semi-spheres are invariant subspaces.  Applying the transformation
(\ref{algebra4}) to $q_{3}$ we get another Lee equation:
\begin{equation}\label{algebra11}
  \frac{dq_{3}^{'}}{dp}=-\frac{\vec{n}\cdot \vec{q}_{\bot}^{'}}{q^2}\ q_{3}^{'}\ \ ,\
  \ q_{3}^{'}(p=0)=q_3~.
\end{equation}
Integrating it we obtain:
\begin{equation}\label{algebra12}
q_{3}^{'}=q_{3} \ \exp \left \{ -\int \frac{\vec{n}\cdot
\vec{q}_{\bot}^{'}}{q^2}dp  \right \}~.
\end{equation}
On the other hand, from (\ref{algebra6}) we can conclude that:
\begin{equation}\label{algebra13}
\vec{n}\cdot \vec{q}_{\bot}^{'}=\frac{  q \sinh(\frac{p}{q}) +
\vec{n}\cdot \vec{q}_{\bot}\cosh(\frac{p}{q})
}{\cosh(\frac{p}{q})+ \frac{\vec{n}\cdot
\vec{q}_{\bot}}{q}\sinh(\frac{p}{q}) }~.
\end{equation}
Substituting this expression to (\ref{algebra12}) and integrating
it we arrive at:
\begin{equation}\label{algebra14}
q_{3}^{'}=q_{3}\ \left (1+\frac{\vec{\kappa }\cdot
\vec{q}_{\bot}}{q^2}\right )^{-1}\left(1-\frac{\kappa
^2}{q^2}\right)^\frac{1}{2}~.
\end{equation}
Combining this expression and relation (\ref{q+x}), finally we
get:
\begin{equation}\label{algebra15}
  \frac{q_{3}^{'}}{\sqrt{q^2 -q_{\bot}^{'2}}}=\frac{q_{3}}{\sqrt{q^2
  -q_{\bot}^{2}}}=\frac{q_3}{|q_3|}=\textrm{inv}~.
\end{equation}
The consequence of this invariance is appearance of the signature
in amplitudes (cross-sections)  which corresponds to creating of
the particle $C$ to the forward or backward semi-sphere.

At the end of this section we note that nonlinear transformations
(\ref{q+x}) can be linearized in the three-dimension space on
hyperboloid. Indeed, if we introduce new variables $u_0 ,u_1,
u_2$:
\begin{equation}\label{algebra16}
  u=(u_0 ,u_1 ,u_2)\ \ ,\  u_0 =\frac{q}{q_3} \ ,\ \vec{u} =\frac{\vec{q}_{\bot}}{q_3}\ \ ,\ \ u^2=u_0 ^2-u_1 ^2-u_2
  ^2=1~,
\end{equation}
than the transformations (\ref{q+x}) can be written in these
variables as:
\begin{equation}\label{algebra17}
  u^{'}=\Lambda (\vec{\kappa })u~,
\end{equation}
with the transformation matrix $\Lambda (\vec{\kappa })$:
\begin{equation}\label{Lambda_d}
\Lambda (\vec{\kappa})=\frac{1}{B} \left(\begin{array}{ccc}
  1                     & \displaystyle\frac{\kappa_1}{q}                    & \displaystyle\frac{\kappa_2}{q}
  \\ \\
  \displaystyle\frac{\kappa_1}{q}    & B+\displaystyle\frac{1-B}{\kappa^2}\kappa_1^2      & \displaystyle\frac{1-B}{\kappa^2}\kappa_1\kappa_2
  \\ \\
  \displaystyle\frac{\kappa_2}{q}    & \displaystyle\frac{1-B}{\kappa^2}\kappa_1\kappa_2  & B+\displaystyle\frac{1-B}{\kappa^2}\kappa_2^2
\end{array} \right)
\end{equation}

$$     {\rm Det}(\Lambda (\vec{\kappa})) =1\ \ ,\ \ B=\left ( 1-\frac{\kappa ^2}{q^2}   \right )^{\frac{1}{2}}~.$$

\section{Representation of $SO_{\mu}(2,1)$ algebra in the transverse momentum
space}

Let us obtain group basic functions, which realize states with a
define value of the Casimir operator (\ref{algebra3}) and definite
value of the third projection of orbital momentum $L_3$. For that
we introduce spherical coordinates in the momentum space:
$$q_1=q\sin\theta\cos\varphi~,$$
$$q_2=q\sin\theta\sin\varphi~,$$
$$q_3=q\cos\theta~.$$
Using an explicit  expression for operators $d_i\ ,\ L_3 $, we
obtain:
\begin{eqnarray}\label{oper_sphere}
d_1 &=&
\frac{i}{q}\Big(\cos\varphi\cos\theta\frac{\partial}{\partial\theta}
-\frac{\sin\varphi}{\sin\theta}\frac{\partial}{\partial\varphi} -\sin\theta\cos\varphi\Big)~,\\
\nonumber d_2 &=&
\frac{i}{q}\Big(\sin\varphi\cos\theta\frac{\partial}{\partial\theta}+
\frac{\cos\varphi}{\sin\theta}\frac{\partial}{\partial\varphi} -\sin\theta\sin\varphi\Big)~,\\
\nonumber L_3 &=& -i\frac{\partial}{\partial\varphi}~.
\end{eqnarray}
The Casimir operator is:
\begin{equation}\label{Ksphere}
  \hat{K}=\frac{-1}{q^2}\Bigg(\cos^2\theta\frac{\partial^2}{\partial\theta^2}
  +\Big(\frac{\cos\theta}{\sin\theta}-3\cos\theta\sin\theta\Big)\frac{\partial}{\partial\theta}
  +\frac{\cos^2\theta}{\sin^2\theta}\frac{\partial^2}{\partial\varphi^2}-2\cos^2\theta\Bigg)~.
\end{equation}
Let us consider a system of equations:
\begin{equation}\label{algebra18}
\hat{K}\Psi = b^2 \Psi \ \ ,\ \ L_3 \Psi = m \Psi~.
\end{equation}
To make analysis simpler we consider first special case of $m=0$.
For this case we have only one ordinary differential equation:
\begin{equation}\label{algebra19}
  (1-x^2)^2 \ \frac{d^2 f}{d x^2}+ \frac{(1-x^2)^2 }{x}\frac{d f}{d x}+ 4 q^2 b^2 f
  =0~,
\end{equation}
 with
\begin{equation}\label{20}
   \Psi (\theta )=\frac{1}{\cos \theta }f(\theta )   ,\ \ x=\tan \frac{\theta
   }{2}~.
\end{equation}
This equation can be reduced to the Riemann equation. The point
$\theta = \pi /2 $ is the singular point of equation
(\ref{algebra19}).  Regular and continuous solution for the region
$ 0\leq \theta < \pi /2 $ is the so called cones function:
 \begin{equation}\label{algebra21}
\begin{split}
\Psi (\theta )=& \frac{1}{\cos \theta}P_{-1/2+i\mu }\left (
\frac{1}{\cos \theta} \right )\ \ ,\
\ 0\leq \theta < \pi /2 \\
& \mu =\left ( q^2 b^2 -1/4  \right )^{1/2} \ \ ,\ \ q^2 b^2 \geq
1/4~,
\end{split}
\end{equation}
with $P_{\nu }(z)$ -- the Legendre functions.

To find the solution in the backward semi-sphere $ \pi/2 \leq
\theta < \pi  $, we note that equation (\ref{algebra19}) is
invariant under the transformation $x \rightarrow 1/x $, which
corresponds to transformation $\cos \theta \rightarrow - \cos
\theta $. So, regular and continuous solution in the region
$\frac{\pi }{2} < \theta \leq \pi $ has the following form:
\begin{equation}\label{algebra22}
  \Psi (\theta )= -\frac{1}{\cos \theta}P_{-1/2+i\mu }\left ( -\frac{1}{\cos \theta} \right )\ \ ,\
\ \frac{\pi }{2} < \theta \leq \pi~.
\end{equation}
Finally, combining (\ref{algebra21}) and (\ref{algebra22}) we
obtain:
\begin{equation}\label{algebra23}
  \Psi (\vec{q}_{\bot} )= \frac{q}{\sqrt{q^2 -q_{\bot}^2}}P_{-1/2+i\mu }\left ( \frac{q}{\sqrt{q^2 -q_{\bot}^2}} \right )\ \ ,\
\ q_{\bot}=q \sin \theta \ \ ,     0 \leq \theta \leq \pi~.
\end{equation}

There is an important physical constraint on the spectrum of the
operator $\hat{K} $:

\begin{equation}\label{algebra24}
  b^2 \geq \frac{1}{4 q^2}~.
\end{equation}
This constraint follows from the Heisenberg uncertainty principle
-- a particle with fixed fixed energy can not be created in the
region of the coordinate transverse plane of
 arbitrary small
radius.

The generalization for case $m\neq 0$ is simple, the solution of
Eqs.~(\ref{algebra18}) has the form:
\begin{equation}\label{algebra25}
\begin{split}
\Psi (\theta , \varphi )=& \frac{1}{|\cos \theta |}P_{-1/2+i\mu
}^{|m|}\left ( \frac{1}{|\cos \theta |} \right ) e^{im\varphi }\ \
,\
\ 0\leq \theta \leq \pi   \ \ ,\ \ 0\leq \varphi \leq 2 \pi~,   \\
& \mu =\left ( q^2 b^2 -1/4  \right )^{1/2} \ \ ,\ \ q^2 b^2 \geq
1/4\ \ ,\ \ m=0,\pm 1,\pm 2,...~.
\end{split}
\end{equation}
The set of functions (\ref{algebra25}) forms complete and
normalizable  basis. The decomposition of functions in this basis
is known as Fock-Meller expansion and for case $ m=0 $ it was
investigated by V.A.Fock \cite{Fok}.

For the scattering problem, we are interested in, only the plane
wave on $SO_\mu(2,1)$ group are relevant. These plane waves
correspond to the main series of irreducible representations on
the group function space, this series is generated by $ \oplus $
operation (\ref{q+x}). The situation is similar to a problem of
building of a relativistic configurational space on Lorentz group
$SO(3,1)$. In our case the role of coordinate operators is played
by motion generators in the momentum space. The momentum space is
represented on positive part of the two-sheet hyperboloid $E^2
-q^2 =m^2$. This problem was solved in
Refs.~\cite{Kadyshevskii,Vilenkin1}. It was shown, that the
relativistic configurational space conjugate to the momentum space
in Fourie-transformation mean. In our case the role of the unitary
transformation kernel from $x$ to $q$ -space  is played by the
plain waves on hyperboloid, so called Shapiro's functions
\cite{Shapiro}. These functions implements the main series of
unitary representation $SO(3,1)$ group.

We do not give detailed calculations, we just discuss main step.
Let the state $\psi (\vec{q}_{\bot})$ diagonalizes an operator
$(\vec{n}\vec{d}_{\bot})$ , with $\vec{n} $ is a vector
perpendicular to the momentum $\vec{p}_{A}$:
\begin{equation}\label{algebra26}
 ( \vec{n}\vec{d}_{\bot}) \ \psi (\vec{q}_{\bot}) = \textrm{const} \ \psi
 (\vec{q}_{\bot})~.
\end{equation}
Now, using the transformation (\ref{q+x}) we obtain that function
\begin{equation}\label{algebra27}
  f(\vec{q}_{\bot})=\sqrt{1-\frac{q^2_ {\bot}}{q^2}}\ \psi (\vec{q}_{\bot})
\end{equation}
satisfies the following functional equation:
\begin{equation}\label{algebra28}
  f(\vec{q}_{\bot}\oplus \vec{\kappa})\ =f(\vec{q}_{\bot})\cdot
  f(\vec{\kappa})~.
\end{equation}
Usual plain waves $\exp \left( i\vec{q}_{\bot}\cdot \vec{b}
\right) $ correspond to translations of the Euclidean plain for
the case when $ \oplus $ operation is just a sum operation.

Obviously, the relation (\ref{algebra28}) is not satisfied for
arbitrary $\vec{\kappa}$ , but only for lines of a coordinate grid
in the momentum space. Let $\kappa _{i}=\kappa _{i} (t) $ be a
parametric representation of those lines. We can assume without
loss of generality that:
\begin{equation}\label{algebra29}
  \kappa _{i} (t=0)=0 \ \ ,\ \ \left ( \frac{d\kappa _{i}}{dt} \right )_{t=0} =n_{i}\ \ ,\ \ n^2=1\ \ ,\ \
  i=1,2~.
\end{equation}
Expanding the equation (\ref{algebra28}) in the vicinity of the
point $\kappa _{i}=0 $ and using that:
\begin{equation}\label{algebra30}
  \left ( \frac{dq_{i}^{'}}{d\kappa _{k}}  \right )_{\vec{\kappa }=0} =\delta
  _{ik}-\frac{q_{i}q_{k}}{q^2}~,
\end{equation}
we obtain the following equation for $f(\vec{q}_{\bot})$:
\begin{equation}\label{algebra31}
 (\delta _{ik} -x_{i}x_{k} )n_{k} \frac{\partial f}{\partial x_{i}}=2i\beta
  f~,
\end{equation}
with $$ x_{i}=\frac{q_{i}}{q} \ \ ,\ \ \left ( \frac{\partial
f}{x_{k}} \right )_{\vec{x}=0}=-i\alpha _{k}\ \ ,\ \ \beta
=\frac{1}{2}(\vec{\alpha }\cdot \vec{n} )\ \ ,\ \ i,k=1,2~.
$$
It is a differential equation in partial derivative of the first
order. Its solution is determined up to arbitrary function of the
scalar $\zeta $ defined as:
$$ \zeta =\frac{1-x^2}{(1-z)^2} \ \ ,\  z=\frac{\vec{n}\cdot \vec{q}_{\bot}}{q}~.$$
The constant $\beta$ is arbitrary as well. This arbitrariness is
fixed by the condition that the function $\psi (\vec{q}_{\bot})$
must be the eigenfunction of the Casimir operator $\hat{K}$:
\begin{equation}\label{algebra32}
  \hat{K}\psi (\vec{q}_{\bot})=b^2 \psi (\vec{q}_{\bot})~.
\end{equation}
Eventually this gives the corresponding equation for the function
$f$:
\begin{equation}\label{algebra33}
  (x_{i}x_{k}-\delta _{ik} )\frac{\partial ^2 f}{\partial x_{i} \partial  x_{k}} +
  2\vec{x}\cdot \frac{\partial f}{\partial \vec{x}} - \frac{b^2 q^2  }{1-x^2} f
  =0~.
\end{equation}
The system of equations (\ref{algebra31}) and (\ref{algebra33}) is
equivalent to the system (\ref{algebra26}) and (\ref{algebra32}).
Its solution gives us the basis functions of plane waves type. Let
us introduce the notation:
$$\psi (\vec{q}_{\bot}) \equiv  \xi
(\vec{q_\bot} ,\vec  \mu)~,$$ with  $\vec{n}$ - the two-dimension
elementary vector, introduced in equation (\ref{algebra26}),
$\vec{\mu} = \vec{n} \mu  $. ~ $\mu $ was defined in
(\ref{algebra21}). The final result is the following:
\begin{equation}\label{algebra34}
\begin{split}
   \xi (\vec{q_\bot} ,\vec  \mu) &=  \frac{q}{\sqrt{q^2-q^2_\bot
}}\left(\frac{q-\vec n \cdot \vec{q_\bot }}{\sqrt{q^2-q^2_\bot }}\right)^{-\frac{1}{2}+i\mu}\\
&=u_0(u\cdot n)^{-\frac{1}{2}+i\mu }\ \ ;\ \ (u\cdot n)=u_0 - \vec{u}\cdot \vec{n}\\
&=u_{0}\left [ u_0 -\sqrt{u_0 ^2 -1}\cos (\varphi -\theta ) \right ]^{-\frac{1}{2}+i\mu }~,\\
  & \vec{q_\bot}=(q_{\bot}\cos \varphi \ , \ q_{\bot}\sin \varphi  )\ \ ,\ \ \vec{n}=(\cos \theta , \sin \theta
  )~,\\
  & n=(1,\vec{n})\ \ ,\ \ n^2=n_{0}^2- \vec{n}^{\ 2}=0~,
 \end{split}
\end{equation}
with $n$ - a three dimensional light-like vector on hyperboloid.

From the relation (\ref{algebra27}) between $f$  and $\psi$ it
follows that:
\begin{equation}\label{35}
  f\equiv f(\vec{q_\bot} ,\vec  \mu)=(u\cdot n)^{-\frac{1}{2}+i\mu
  }~.
\end{equation}
This is the two dimensional Shapiro function \cite{Shapiro}. Its
form has universal form, because the irreducible representation of
$SO(m,1)$ group is realized by the same function only with a
dimension of hyperboloid vectors $n$ and $u$ equals to $m+1$.

The function $f(\vec{q_\bot} ,\vec  \mu)$ has simple asymptotic
expansion for $ (q_{\bot}/q ) \ll 1 \ \ , \ (1/q^2 b^2)\ll 1 $:
\begin{equation}\label{36}
  f(\vec{q_\bot} ,\vec  \mu)=e^{i\vec{b}\cdot \vec{q}_{\bot}} \left ( 1 +
  O\left(\frac{q_{\bot}}{q}\right) + O\left(\frac{1}{q^2 b^2}\right) \right )
  \ \ ,\ \ \vec{b}=\vec{n}b~.
\end{equation}
It shows us a link between plain waves on $SO_\mu(2,1)$ group and
its analogon on $E(2)$ group. We see that the parameter $\vec b$
corresponds to the "impact parameter" in the infinite momentum
frame, e.g. \cite{Budkard,Diehl}.  The formulae we derive
generalize the "impact parameter" to arbitrary frame and can be
used in whole range  $q^2 b^2 \geq 1/4$.  Generically, the
equations for arbitrary frame can be obtain from the equations in
the infinite momentum frame by the change the transformation
kernel:
$$\exp(i \vec{q}_{\bot} \cdot \vec{b})\ \rightarrow \xi (\vec{q}_{\bot} , \vec{\mu })~,$$
and one has to introduce  $|\vec{\mu},q  \rangle _{SO_{\mu}(2.1)}$
state instead $|\vec{b},q^+  \rangle _{E(2)}$. Also these changes
solve the problem of direct and reverse transformations
$(\vec{q}_{\bot})\rightleftarrows (\vec{b}) $, which is connected
with a finiteness of the range for $q_{\bot}$ at fixed $E_{q}$  or
$p^+ $.

Here we show one more relation, usual for plane waves on the
Euclidean plane and on the surface defined by (\ref{q+x}). As it
is shown in \cite{Beitman}:
\begin{equation}\label{algebra39}
 \int \limits _{0}^{2\pi} d \theta \ f(\vec{q_\bot} ,\vec  \mu)=2\pi P_{-1/2
 + \mathrm{i}\mu}(\frac{q}{\sqrt{q^2-q^2_\bot }})~,
\end{equation}
with  $\theta $ - direction angle of vector $\vec{n}$. Now we use
Fock expansion \cite{Fok}:
\begin{equation}\label{algebra40}
   P_{-1/2 + \mathrm{i}\mu}(\cosh \alpha )= \sqrt{\frac{\alpha }{\sinh \alpha}}\left \{ J_{0}(\mu \alpha
   )+\frac{1}{8 \mu }\left ( \coth \alpha  - \frac{1}{\alpha } \right )J_{1}(\mu \alpha )+ ......
   \right \}~.
\end{equation}
Substituting (\ref{36})  and  (\ref{algebra40}) to equality
(\ref{algebra39}) and taking limit $(q_{\bot}/q ) \ll 1 \ \ , \
(1/q^2 b^2)\ll 1 $, we get a relation well-known from the eikonal
formalism
\begin{equation}\label{algebra41}
  \int \limits _{0}^{2\pi} d \theta \ e^{i\vec{b}\cdot \vec{q}_{\bot} }= 2\pi J_{0}(b
  q_{\bot})~.
\end{equation}

A system of functions $\xi (\vec{q_\bot} ,\vec \mu) $ form a
complete orthogonal system independently on the forward and the
backward semi-spheres
 ($q_3 = \pm\sqrt{q^2 - q_{\bot }^{2}}$):
\begin{equation}\label{algebra37}
  \frac{1}{(2\pi)^2} \int\xi (\vec{q_\bot} ,\vec  \mu)\ \bar{\xi }(\vec{q_\bot} ,\vec { \mu
  }')\ \mathrm{d}\Omega_{\vec{q}}=\frac{1}{\tanh(\pi\mu)}\ \delta^{(2)}(\vec{\mu}-\vec{\mu
  }')\quad\textrm{--orthogonality}~,
\end{equation}

\begin{equation}\label{algebra338}
 \frac{1}{(2\pi)^2} \int\xi (\vec{q_\bot} ,\vec  \mu)\ \bar{\xi }(\vec{q_\bot}' ,\vec  \mu
  )\ \mathrm{d}\Omega_{\vec{\mu}}=q(q^2-q^2_\bot )^\frac{1}{2}\
  \delta^{(2)}(\vec{q_\bot}-\vec{q_\bot}')\quad\textrm{--compliteness}~,
\end{equation}
with:
\begin{equation}\label{7}
  \mathrm{d}\Omega_{\vec{q}}=\frac{1}{q\sqrt{q^2-q^2_\bot }}\ \mathrm{d}\vec{q_\bot}\ ,\quad
  \mathrm{d}\Omega_{\vec{\mu}}=\tanh (\pi\mu)\ \mathrm{d}\vec{\mu}~,
\end{equation}
$$\quad\vec{\mu}=(\mu \cos\theta ,\mu \sin\theta )\ ,\ \
 \delta^{2}(\vec{\mu}-\vec{\mu}')=\frac{1}{\mu}\ \delta(\mu-\mu')\ \delta(\theta-\theta')~.$$
Any function $F(\vec{q})$ can be decomposed unambiguously in the
$\xi (\vec{q_\bot} ,\vec  \mu)$ basis for forward semi-sphere $q_3
> 0 $,  as a $SO_{\mu}(2,1)$ group
 function (or independently for $ q_3  < 0 $):
\begin{equation}\label{algebra42}
\begin{split}
  F(\vec{q})=\int \xi (\vec{q_\bot} ,\vec  \mu)  \ C(\vec{\mu})\ \mathrm{d}\Omega_{\vec{\mu}}~,\\
C(\vec{\mu})=\frac{1}{(2\pi)^2} \int\bar{\xi} (\vec{q_\bot} ,\vec
\mu) F(\vec{q})\ \mathrm{d}\Omega_{\vec{q}}~.
 \end{split}
\end{equation}
These relations allow us to build N-particles Fock space, where
one particle in state with the fixed spatial parameter
$\vec{\mu}$.

\section{Fock space on $SO_{\mu}(2.1)$ group}
Now let us consider the process, where one of the created
particles is in the state with a definite spatial  parameter
$\vec{\mu}$ and with a definite value of energy  $\mathrm{E_{q}}$
and with a definite sign of projection of $z$-momentum. We denote
it as:
\begin{equation}\label{Foc2}
  |\vec{\mu},q,\epsilon\rangle \ ;\ \quad \epsilon=\pm 1\ \textrm{--sign of
projection}~.
\end{equation}

In the same way we  write state with a definite value of the
transverse momentum $\vec{q}_{\bot}$, with a definite value of
energy $\mathrm{E_{q}}$, with a definite sign of projection of the
$z$-momentum. We denote it as :
\begin{equation}\label{Foc3}
 |\vec{q}\rangle  ^{\pm }= |\vec{q}_{\bot},q_{3}=\pm \sqrt{q^2-q_{\bot}^2}\rangle
\end{equation}
with
\begin{equation}\label{Foc5}
 |\vec{q}\rangle  = a^{+}(\vec{q}) |0\rangle  \ \ ,\ \ [ a(\vec{q}) \  a^{+}(\vec{p}) ]= \delta (
\vec{q}-\vec{p} )~.
\end{equation}

These two types of states are related to each other by the
following transformation  on the sphere $q^2=\mathrm{const}$, see
eqs.~(\ref{Foc2}) and (\ref{Foc3}):
\begin{equation}\label{Foc4}
  \langle \vec{q}_{\bot},\epsilon \sqrt{q^2-q_{\bot}^2}|=\int \xi (\vec{q_\bot} ,\vec  \mu)
  \langle \vec{\mu},q,\epsilon|\ \mathrm{d}\Omega_{\vec{\mu}}~,
\end{equation}
$$\langle \vec{\mu},q,\epsilon|=\frac{1}{(2\pi)^2}
\int\bar{\xi} (\vec{q_\bot} ,\vec  \mu)\langle
\vec{q}_{\bot},\epsilon \sqrt{q^2-q_{\bot}^2}|\
\mathrm{d}\Omega_{\vec{q}}~.$$
 It is
consequence of completeness  and orthogonal properties of
functions $\xi (\vec{q_\bot} ,\vec  \mu)$.

The integral relation is correct for a arbitrary function
$F(\vec{q})$:
\begin{equation}\label{Foc1}
  \int F(\vec{q})\  \mathrm{d}\vec{q}= \int q^2 dq \
  \mathrm{d}\Omega \ F(\vec{q})=\sum\limits _{\epsilon=\pm 1}\int q^2 dq \
  \mathrm{d}\Omega_{\vec{q}}\ F(\vec{q}_{\bot},\epsilon
  \sqrt{q^2-q_{\bot}^2})~,
\end{equation}
$$\mathrm{d}\Omega =sin \theta d \theta d\varphi \ \ , \ \mathrm{d}\Omega_{\vec{q}} = \frac{1}{q\sqrt{q^2-q^2_\bot }}\ \mathrm{d}\vec{q_\bot}~. $$
Using this relation and representation (\ref{Foc4}) we obtain for
the unit operator in the one- particle Fock space:
\begin{equation}\label{Foc6}
  \begin{split}
    \hat{\mathrm{I}}&=\int|\vec{q}\rangle \langle \vec{q}|\ \mathrm{d}\vec{q} =\\
             &=\sum \limits_{\epsilon=\pm 1}\int q^2 dq \ \mathrm{d}\Omega_{\vec{q}}\ |\vec{q}_{\bot},\epsilon \sqrt{q^2-q_{\bot}^2}\rangle \langle \vec{q}_{\bot},\epsilon \sqrt{q^2-q_{\bot}^2}| =\\
             &=(2\pi)^2 \sum \limits_{\epsilon=\pm 1}\int q^2 dq \ \mathrm{d}\Omega_{\vec{\mu}}\
             |\vec{\mu},q,\epsilon\rangle \langle \vec{\mu},q,\epsilon|~.
    \end{split}
\end{equation}
The last relation defines the matrix element $\langle
\vec{\mu},q,\epsilon|\vec{\nu},k,\eta\rangle $  unambiguously.
Indeed we obtain:
\begin{eqnarray}\label{Foc7}
|\vec{\nu},k,\eta\rangle
&=&\hat{\mathrm{I}}|\vec{\nu},k,\eta\rangle = (2\pi)^2 \sum
\limits_{\epsilon=\pm 1}\int q^2 dq \ \mathrm{d}\Omega_{\vec{\mu}}
             |\vec{\mu},q,\epsilon\rangle \langle \vec{\mu},q,\epsilon|\vec{\nu},k,\eta\rangle
             \\\nonumber
 &=&\sum \limits_{\epsilon=\pm 1} \int \left \{(2\pi)^2 q^2  \tanh (\pi\mu)\
 \langle \vec{\mu},q,\epsilon|\vec{\nu},k,\eta\rangle   \right\}|\vec{\mu},q,\epsilon\rangle  dq\
 \mathrm{d}\vec{\mu}~.
\end{eqnarray}
It follows that
\begin{equation}\label{Foc8}
(2\pi)^2 q^2 \tanh (\pi\mu)
 \langle \vec{\mu},q,\epsilon|\vec{\nu},k,\eta\rangle =\delta ^{(2)}(\vec{\mu}-\vec{\nu})\  \delta (q-k) \ \delta _{\epsilon \eta}
\end{equation}
and the expression for matrix element of conversion is:
\begin{equation}\label{Foc9}
  \begin{split}
 & \langle \vec{q_{\bot}},\epsilon\sqrt{q^2-q^2_{\bot}}\ |\vec{\nu}
  ,k,\eta \ \rangle  = \\
  &=\int\xi(\vec{q}_{\bot}, \vec{\mu})\ \langle \vec{\mu},q,\epsilon \
  |\vec{\nu},k,\eta\rangle  \ \mathrm{d}\Omega_{\vec{\mu}}=\\
  &=\xi(\vec{q}_{\bot},\vec{\nu})\frac{1}{(2\pi)^{2}}\ \delta_{\epsilon \eta}\ \frac{1}{q^{2}}\
  \delta(k-q)~.
  \end{split}
\end{equation}
Here we used eq.~(\ref{Foc4}). We  obtain from the condition
$\langle \vec{q}|\vec{k}\rangle =\delta^{(3)}(\vec{q}-\vec{k})$
that:
\begin{equation}\label{Foc10}
 \begin{split}
  &\langle \vec{q}_{\bot} ,\epsilon\sqrt{q^2-q_{\bot}^2}|\vec{k}_{\bot} ,\eta\sqrt{k^2-k_{\bot}^2}\rangle =\\
  &=\delta ^{(2)}(\vec{q}_{\bot}-\vec{k}_{\bot})\ \delta
  (\epsilon\sqrt{q^2-q^2_{\bot}}-\eta\sqrt{k^2-k^2_{\bot}})=\\
  &=\delta _{\epsilon \eta}\ \delta ^{(2)}(\vec{q}_{\bot}-\vec{k}_{\bot})\ \delta
  (q-k)\frac{\sqrt{k^2-k_{\bot}^2}}{k}~.
 \end{split}
\end{equation}
To calculate matrix elements (\ref{Foc8}), (\ref{Foc9}) and (
\ref{Foc10}) on the plane $q=k$
 it is necessary to define the expression for $\delta (q-k)$. Let us use the well-known
 relation:
\begin{equation}\label{Foc11}
  \delta (q-k)=\frac{k}{E_{k}}\delta (E_{q}-E_{k})~.
\end{equation}
This implies:
\begin{equation}\label{Foc12}
   \delta (q-k)=\frac{k}{E_{k}}\delta (E_{q}-E_{k})=\frac{k}{E_{k}}\frac{T}{2\pi}=\frac{|\vec{V}|\cdot
   T}{2\pi}~,
\end{equation}
\begin{equation}\label{Foc13}
  \frac{\sqrt{k^2-k_{\bot}^2}}{k}\cdot\delta (q-k)=\frac{|k_3|}{E_{k}}\delta (E_{q}-E_{k})=\frac{|V_{3}|\cdot
  T}{2\pi}~,
\end{equation}
with $\vec{V}=\frac{\vec{k}}{E_{k}}$ -- a velocity of particle
with momentum $\vec{k}$\ ,\ and   $T$ - the infinite interval of
time.

Let us consider the process with $s$ different particles in the
finite state with momenta  $\vec{q_1},\vec{q_2},\ldots \vec{q_s}$
and one particle in the state $|\vec{\mu},q,\epsilon\rangle $. The
basis of vectors in $(s+1)$ -particles Fock space are the direct
product of one-particle basic vectors:
\begin{equation}\label{Foc14}
  |\{ q_s \} ; \vec{q}\rangle  \equiv |\vec{q_1},\vec{q_2},\ldots \vec{q_s} ;\vec{q}
  \rangle ~,
\end{equation}
\begin{equation}\label{Foc15}
  |\{ q_s \} ;\vec{q}_{\bot},\epsilon \sqrt{q^2-q_{\bot}^2}\rangle  \equiv |\vec{q_1},\vec{q_2},\ldots \vec{q_s};\vec{q}_{\bot},\epsilon
  \sqrt{q^2-q_{\bot}^2}\rangle ~,
\end{equation}
\begin{equation}\label{Foc16}
 |\{ q_s \};\vec{\mu},q, \epsilon\rangle  \equiv |\vec{q_1},\vec{q_2},\ldots
 \vec{q_s};\vec{\mu},q,\epsilon\rangle ~.
\end{equation}
These are equivalent complete sets of basic vectors  in the
$(s+1)$ -particles Fock space, with $s=0,1,....$ . Hence there are
three representation  of the unit operator in the Fock space.
\begin{equation}\label{Foc17}
  \begin{split}
    \hat{\mathrm{I}}&= \sum \limits_{s}\int (\prod \limits^{s}_{i=1}\mathrm{d}\vec{q_i})\ \mathrm{d}\vec{q}\ |\{ q_s \};\vec{q}\rangle \langle \{ q_s \};\vec{q}|\\
&=\sum \limits_{s}\sum \limits_{\epsilon=\pm 1}\int (\prod
\limits^{s}_{i=1}\mathrm{d}\vec{q_i})\ q^2 dq\
\mathrm{d}\Omega_{\vec{q}}\ |\{ q_s \};\vec{q}_{\bot},\epsilon \sqrt{q^2-q_{\bot}^2}\rangle \langle \{ q_s \};\vec{q}_{\bot},\epsilon \sqrt{q^2-q_{\bot}^2}| \\
    &=(2\pi)^2\sum \limits_{s}\sum \limits_{\epsilon=\pm 1}\int (\prod \limits^{s}_{i=1}\mathrm{d}\vec{q_i})\ q^2 dq \
    \mathrm{d}\Omega_{\vec{\mu}}\ |\{ q_s \};\vec{\mu},q,\epsilon\rangle \langle \{ q_s
    \};\vec{\mu},q,\epsilon|~.
  \end{split}
\end{equation}
From last expression it is follows that
$$ \langle \{ k_s
\};\vec{\nu},k,\eta|\{ q_s \};\vec{\mu},q,\epsilon\rangle = $$
 \begin{equation}\label{Foc18}
(2\pi)^2 q^2 \tanh (\pi\mu)
 \langle \{ q_s \};\vec{\mu},q,\epsilon|\{ k_s \};\vec{\nu},k,\eta\rangle =\delta ^{(2)}(\vec{\mu}-\vec{\nu})
 \  \delta (q-k) \ \delta _{\epsilon \eta}\ \prod \limits^{s}_{i=1}\ \delta ^{(3)}(\vec{q}_i -
 \vec{k}_i)~.
\end{equation}

This matrix element appears in a calculation of  a cross section
and will be used in next section.

\section{Cross sections with creation of particle in a state from $SO_{\mu}(2.1)$ }

Let us obtain a expression for cross-section through matrix
elements of S-matrix, with final state defines by a vector of Fock
space $\langle \{ q_s \};\vec{\mu},q,\epsilon|$ , i.e. with $s+1$
created particles, and one of them has a state from
$SO_{\mu}(2,1)$. Formalism of such procedure describes in details
in \cite{Shirkov}. We note only  some important points. Basics of
formalism is a quantum-mechanic interpretation of norm
one-particle state $|\psi \rangle $. This state can be represented
as wave packet:
\begin{equation}\label{s1}
  |\Psi \rangle  =\int f(\vec{k})\ a^+(\vec{k})\ \mathrm{d}\vec{k}|0\rangle \ ,\quad [a(\vec{k})\ a^+(\vec{q})] =
             \delta ^{(3)}(\vec{k}-\vec{q})
\end{equation}
with $f(\vec{k})$ - wave function, which gives a quantum-mechanic
description of state $|\Psi\rangle $. Here, if the norm of that
state
$$ ||\Psi \rangle |^2 =\langle \Psi|\Psi \rangle =\int |f(\vec{k})|^2\
\mathrm{d}\vec{k}
$$
equals to one, then $|f(\vec{k})|^2$ is interpreted as
corresponding probability density (Born interpretation). If the
norm is equal to $N>1$, then $ f(\vec{k})$ is interpreted as an
one-particle distribution function  of particles in statistical
assembly with N particles.

A different type particles state is constructed as direct product
of corresponding set of states. A norm of that state equals to
product of a number of particles in the set:
$$ ||\Psi \rangle |^2=N_1\cdot N_2 \cdot  \cdots N_s $$

As an example, let us consider a set of particles with fixed
momentum $\vec{k}$:
$$ |1\rangle =(2\pi)^{3/2}n^{1/2}\ a^+(\vec{k})|0\rangle $$ A sense of parameter $n$ follows from a definition of norm:
$$||1 \rangle |^2=(2\pi)^{3}n\ \delta (0)=n\cdot V =N$$
So, $n=\frac{N}{V}, $ i.e. it is a density of particles.

The wave function $f(\vec{k})$ defined in (\ref{s1}) for such flux
is
$$f(\vec{k})=(2\pi)^{3/2}n^{1/2}\ \delta(\vec{k}-\vec{k}_0) $$
So the state $|1\rangle $ is the state with infinite norm and
describes flux of particles with fixed momentum $\vec{k}$ and
density of particles in the flux $n$. If we have $s$ particle
fluxes with different momenta
$\vec{q}_1,\vec{q}_2,\cdots,\vec{q}_s$, then corresponding state
is defined as direct product of states for every flux:
\begin{equation}\label{s2}
  |s\rangle =\prod \limits^{s}_{i=1}\left \{(2\pi)^{3/2}n_{i}^{1/2}\ a^+(\vec{q}_{i})      \right\}|0\rangle
\end{equation}
A norm of such state is
\begin{equation}\label{s3}
  ||s \rangle |^2=N_1\cdot N_2 \cdot  \cdots N_s
\end{equation}
with $$ N_i=n_iV\ ,\ i=1,2,\ldots s  $$

When we consider a collision of two beams with defined momenta and
density, we build an initial state $|in\rangle $ as  (\ref{s2})
state with $s=2$:
\begin{equation}\label{s4}
  |in\rangle =|s=2\rangle ~.
\end{equation}
 We may interpret such product at $s=2$ in context of a
dynamical chaos hypothesis, as an average collisions number of
particles from different beams. "Star" of finite state particles
appears in every such collision. By virtue of the unitary relation
$SS^+ = I$ we have:
\begin{equation}\label{unit}
  \left | |out \rangle  \right |^2 = \left | |in \rangle  \right |^2
\end{equation}
with $|\ out\rangle =\mathrm{\hat{S}}|\ in\rangle  $~. So, the
norm of $|out \rangle $ state defines a number of "stars" burned
during infinity time $T$ and in infinity volume $V$.

We are interested in process $2\rightarrow s+1 \ \ ,\ s=0 ,1,...$,
so let us decompose $|\ out\rangle $ state on  (\ref{Foc15} ,
\ref{Foc16}) basis.  We have:
\begin{eqnarray}\label{s5}
|out\rangle &=&~\\\nonumber
 &=&(2\pi)^2\sum \limits_{\epsilon=\pm 1}\int
(\prod \limits^{s}_{i=0}\mathrm{d}\vec{q_i})\ q^2 dq \
   \mathrm{d}\Omega_{\vec{\mu}}|\{ q_s \};\vec{\mu},q,\epsilon\rangle \langle \{ q_s
    \};\vec{\mu},q,\epsilon|\hat{S}|in\rangle \\ \nonumber
    &=&\sum \limits_{\epsilon=\pm 1}\int (\prod \limits^{s}_{i=0}\mathrm{d}\vec{q_i})\ q^2 dq
   \mathrm{d}\Omega_{\vec{q}}|\{ q_s \};\vec{q_{\bot}},\epsilon\sqrt{q^2-q^2_{\bot}}\ \rangle \langle \{ q_s
    \};\vec{q_{\bot}},\epsilon\sqrt{q^2-q^2_{\bot}}\
    |\hat{S}|in\rangle .
\end{eqnarray}
Total numbers of events of $(s+1)$-particles creation during
infinity time and in the whole space are defined by the
$|out\rangle $-state norm:
\begin{equation}\label{s6}
\begin{split}
  N&=||out\rangle |^2=|\hat{S}-\hat{I}|in\rangle |^2=\\
  &=(2\pi)^2\sum \limits_{\epsilon=\pm 1}\int (\prod \limits^{s}_{i=1}\mathrm{d}\vec{q_i})\ q^2
dq \
   \mathrm{d}\Omega_{\vec{\mu}} \ |\langle \{ q_s
    \};\vec{\mu},q,\epsilon|(\hat{S}-\hat{I})|in\rangle |^2 \\
  &=\sum \limits_{\epsilon=\pm 1}\int (\prod \limits^{s}_{i=1}\mathrm{d}\vec{q_i})\ q^2
dq \
   \mathrm{d}\Omega_{\vec{q}} \ |\langle \{ q_s
    \};\vec{q_{\bot}},\epsilon\sqrt{q^2-q^2_{\bot}}\
    |(\hat{S}-\hat{I})|in\rangle |^2~.
  \end{split}
\end{equation}
Here we have used an assumption that all $s+1$ particles are
different. So their operators of creation and annihilation do not
correlate between each other. Also we have used expression
(\ref{Foc18}) for the matrix element.

Relations (\ref{s6}) are initial for obtaining differential cross
sections on corresponding variables (part II). Also they give a
general norm on a total events $A+B \rightarrow C + \{ s \}$,
without any corresponding to a finite states.

Cross section for a process $\sigma(1\rightarrow f)$ (a constant
external field) and a process $\sigma(2\rightarrow f)$ follow from
(\ref{s6}). After extracting kinematic factors we have:
\begin{equation}\label{s7}
  \sigma(1\rightarrow f)=\frac{E_p}{p n T}|(S-I)|in\rangle |^2 \ , \
  |in\rangle =(2\pi)^{3/2}n^{1/2}a^{+}(\vec{p})|0\rangle ~,
\end{equation}
$$  \sigma(2 \rightarrow f)= \frac{1}{n_1 n_2 T V |\vec{u}|}| (S-I)|in\rangle |^2 \ ,\
|in\rangle =(2\pi)^{3}n^{1/2}_1
n^{1/2}_2a^{+}_{A}(\vec{p}_1)a^{+}_{B}(\vec{p}_2)|0\rangle  ~,$$
with $\vec{u}$ -- relative velocity of two particles.

\section{Conclusion}
We have shown that algebra of Poincare group generators contains
subalgebra, which represents $SO(2,1)$ algebra on the $q^2=const$
sphere. The generators of this algebra $d_1\ ,\ d_2\ \ ,\ L_3 $
have exact physical interpretation. The quantum numbers of state,
built with help if this algebra, define the coordinates of
particle creation effective area. They form a complete system of
states in the one-particle Fock space. It allows us to
unambiguously connect a total cross section with a corresponded
$S$-matrix element. That allows one to analyze a spatial structure
of hadrons using various processes experimental data. The
application of the developed formalism to various processes will
be presented elsewhere.

\section{Acknowledgements}
This work was supported in part
 by grant of the President of Russian Federation
 for support of leading scientific schools NSh -5362.2006.2
 (O.N.S \& A.N.V),
 by the Deutsche Forschungsgemeinschaft,
 the Heisenberg--Landau Programme grant 2007,
 and the Russian Foundation for Fundamental Research
 grants No.\ 06-02-16215 and 07-02-91557 (A.A.V)

\end{document}